\documentclass[prb,twocolumn,superscriptaddress,showpacs,%
preprintnumbers,amsmath,amssymb]{revtex4}
\usepackage{amsmath}
\usepackage{graphicx}
\usepackage{dcolumn}

\begin{document}
\title{Transport properties of a 2DEG in the presence of a tilted magnetic field}
\author{A.C.A. Ramos}
\email{acaramos@fisica.ufc.br} \affiliation{Departamento de
F\'isica,
Universidade Federal do Cear\'a, Caixa Postal 6030, Campus do Pici, 60455-760 Fortaleza, Cear\'a, Brazil}%
\author{G.A. Farias}
\email{gil@fisica.ufc.br} \affiliation{Departamento de F\'isica,
Universidade Federal do Cear\'a, Caixa Postal 6030, Campus do Pici, 60455-760 Fortaleza, Cear\'a, Brazil}%
\author{R.N. Costa Filho}
\email{rai@fisica.ufc.br} \affiliation{Departamento de F\'isica,
Universidade Federal do Cear\'a, Caixa Postal 6030, Campus do
Pici, 60455-760 Fortaleza, Cear\'a, Brazil}
\affiliation{Consortium of the Americas for Interdisciplinary
Science, University of New Mexico, Albuquerque, New Mexico 87131,
USA}
\author{N.S. Almeida}
\email{nalmeida@fisica.ufc.br} \affiliation{Departamento de F\'isica, Universidade Federal do Cear\'a, Caixa Postal 6030, Campus do Pici, 60455-760 Fortaleza, Cear\'a, Brazil}%
\date{ \today }

\begin{abstract}
The effect of a tilted dc magnetic field on the transport
properties of a two dimensional electron gas (2DEG) is studied. The influence
of the component of magnetic field parallel to the surface is
analyzed and the dependence of the Fermi energy on the
intensity and direction of the external field is discussed. Numerical
results are obtained for the Hall conductivity considering values of
electronic densities and strength of the dc magnetic field that
are currently obtained in many laboratories.
\end{abstract}


\maketitle

\bigskip
\section{\label{section 1}introduction}

The development of new growing techniques enables control of the
potential responsible for confining the two-dimensional electron
gas (2DEG) in a nanometer scale. This potential influences the
electronic transport properties of the 2DEG and is a key feature
that allows us to have systems with profiles appropriate for
different purposes in the development of new electronic devices.
In addition, when a dc magnetic field is applied to a 2DEG it
modifies the effective potential felt by the electrons and can be
responsible for externally controlled modifications of the
electronic properties of the system. It is expected that the
electronic spin affects the transport properties of the system and
a tilted or oscillating magnetic field modifies the state of the
spin as well as its dynamical characteristics. Although the large
number of papers dealing with this subject
\cite{ref1,ref2,ref3,ref4,ref5,ref5a}, we are still far from
achieving a complete understanding of the role of the intrinsic
magnetic moment in the transport phenomena.

It is supposed that the control of the transport properties via
the spin will allow the substitution of the actual generation of
electronics equipments by the spin-based devices \cite{ref5,ref6}
that can be used, for example, for future applications in quantum
computation.

The two-dimensional systems studied here are constituted by free
electrons confined in a plane with a very narrow thickness
compared to its lateral dimensions. In the ``real world" these
systems might be formed by inversion layers of GaAs/AlGaAs
structures where the electrons are affected by a periodic
potential (that may be very weak) as well as spin-orbit
interaction \cite{ref6a,ref7}. These interactions are important in
order to fully describe the system, but their main effect is to
mix the free electron states (they couple the Landau-level
subbands). In other words, these interactions will introduce
corrections but they will not interfere with the main
characteristics of the physical properties studied here.

In this paper we study the effect of a tilted magnetic field on
the electrical transport properties of a free 2DEG. We name the
two-dimensional region where the electrons are confined the $x-y$
plane, and we set the magnetic field in the $x-z$ plane. In
Section II we develop the formalism necessary to describe a
two-dimensional electronic system in presence of a tilted magnetic
field, taking into account the orbital and spin variables. Since
the Fermi energy depends on the energy and degeneracy of the
states and is one of the key quantities to understand the
transport properties of any system, the unique behavior of the
Landau-level subbands with the spin and field configuration
suggests that this quantity should be carefully analyzed. This is
done in Section III where we calculate and discuss the dependence
of this quantity with the strength and direction of the dc
magnetic field. In order to estimate the influence of the combined
effect of the spin and field configuration on the transport
properties of 2DEG, in  Section IV we obtain numerical results for
the Hall conductivity of low electronic density systems. The
conclusions and general comments are presented in Section V.

\section{\label{section 2}formalism: The one particle eigenstates}

A noninteracting two-dimensional electronic system may describe a
moderate or low electronic density system observed at the
interfaces of semiconductors and materials with a wider bandgap
such as InGaAs/InAlAS or AlGaAs/GaAs. In these structures the
electrons are confined in a narrow quantum well that can be
less than $100$ \AA $\,$wide. In these cases we may write the $z$
dependence of the wave function as a Gaussian function of
half-width $\delta$ equal to the thickness of the system.
Roughly speaking $\delta$ is the width of the quantum well that
confines electrons in the $x-y$ plane, and here we take
$\delta=50$ \AA$\,$ for all numerical purposes.

The procedure described above gives a result similar to those
obtained many years ago by several authors \cite{ref9a} for spin
independent solutions of this system. Here we go further and we
analyze the contribution of the spin (via Zeeman interaction). As
will be seen, in presence of tilted magnetic fields the spin not
only split the levels, but also produces a singular mixture of the
Landau-levels subbands with significative consequences on the
transport properties of the system.

 The one electron Hamiltonian for
this system is writen as:
\begin{equation}
H=-\frac{e\hbar}{2\mu_ec}{\vec\sigma}\cdot{\vec
B}+\frac{1}{2\mu_e} \left(\vec{p}-\frac{e}{c}{\vec A}\right)^{2}
 \label{lab01}
\end{equation}
where the first term is the interaction between the particle's
spin and the magnetic field, and ${\vec\sigma}$ denotes the Pauli
matrices. The second term is the electron kinetic energy with
effective mass $\mu_e$ and charge $-e$ in the presence of a dc
magnetic field ${\vec B}= (B_{x},0,B_{z})$. The magnetic field is
just $\nabla\times{\vec A}$, where the vector potential is  ${\vec
A}=(-B_{z}y,-B_{x}z,0)$.

We have assumed the $z$ dependence of the wave functions as a
Gaussian function and then the effective Hamiltonian (the total
Hamiltonian averaged in the $z$ direction) may be written as:
\begin{eqnarray}
H_{\textit{eff}}
&=&\frac{1}{2\mu_e}(p_x^2+p_y^2)-\frac{e\hbar}{2\mu_ec}({\sigma}_xB_x+
{\sigma}_zB_z) \nonumber\\
& &+ \frac{1}{2}\mu_e {\omega_z}^2(y-y')^2
+\frac{e^2B_{x}^2\delta^2}{2\mu_ec^2}, \label{lab02}
\end{eqnarray}
where $\omega_z=eB_z/\mu_ec$. Since $p_{x}= \hbar
k_{x}$ is a constant (it commutes with $H$) and $y'={l_{c}}^2
k_{x}$ (${l_{c}}^2=\hbar c/eB_z$) is also a constant, a little
algebra reduces the search for the one electron eigenstates
$|n,k_{x},\sigma\rangle =|n\rangle|k_{x}\rangle|\sigma\rangle$ to
the search of $|n\rangle$ and $|\sigma\rangle$ that are solutions
of
\begin{equation}
-\frac{e\hbar}{2\mu_ec}({\sigma}_xB_x+ {\sigma}_zB_z)
|\sigma\rangle=E_{\sigma}|\sigma\rangle \label{lab02a}
\end{equation}
and
\begin{equation}
\left(\frac{{p_y}^2}{2\mu_e}+\frac{1}{2}\mu_e\omega_z^2y^2\right)|n\rangle=\left(E_n-\frac{e^2B_{x}^2\delta^2}{2\mu_ec^2}\right)|n\rangle.
\label{lab02b}
\end{equation}

From  Eq. (\ref{lab02a}) we find the eigenstates and eigenenergies
given by $ |\sigma_\pm\rangle  =  \pm \sin(\theta/2)| \mp \rangle+
\cos (\theta/2)|\pm\rangle$,  $E_{\sigma_\pm }=\pm\hbar\omega_c/2$
respectively. Here $|\pm\rangle$ are the eigenstates of $\sigma_z$
($\sigma_{z}|\pm \rangle = \pm |\pm\rangle$ ),
$\omega_c={e}\sqrt{B_x^2  + B_z^2 }/{\mu_ec}$ and $\theta =
tg^{-1}(B_x/B_z)$. The solutions of Eq. (4) are Landau like states
given by:
\begin{equation}
E_n=\left(n+\frac{1}{2}\right)\hbar \omega_c cos(\theta) +
\frac{1}{2}\mu_e\omega_c^2\delta^{2}sin^2(\theta),
\end{equation}
and then, the eigenstate $|k_{x},n,\sigma\rangle$ might be
represented by the eigenfunction:
\begin{multline}
\Psi_{k_x }^{\sigma_\pm  } (\vec r) = \frac{{e^{ik_x x} }}{{\sqrt {L_x } }}\phi _n (y - y')\\
\times \left[ { \pm \sin \left( {\frac{\theta }{2}} \right)| \mp \rangle  + \cos \left( {\frac{\theta }{2}} \right)| \pm \rangle}\right]
\label{lab022c}
\end{multline}
where $L_x$ is the length of the system along the $x$-direction,
and
\begin{equation}
\phi _n (y - y') =\frac{e^{-\frac{1}{2}\left[(y-y')/{l_c}\right]^2}}{\sqrt {\pi ^{1/2} 2^n n!l_c }} {H_n [(y - y')/l_c]}
\end{equation}
is the one-particle harmonic
oscillator function of frequency $\omega_{c} cos (\theta)$
centered at $y'=k_xl_c^2$, and $n$ is the Landau-index state.

In Figure 1 we display the behavior of the energy of the
eigenstates $|n, \sigma \rangle $ for $1 \leq n \leq 10$. The
graphic shows the modification introduced by the parallel
component of the magnetic field. As usual, the Landau-index gives
the information about the orbital behavior of the system, while
the spin (which has its state controlled by the magnetic field) is
responsible for the shift of each level. As can be seen in this
figure, when the parallel component increases, at some point there
is an inversion in the sequence of the energies of the states and
higher Landau levels become lower energy states. This behavior
will give a unique characteristic to these systems since it will
be responsible for an externally controlled modification of the
Fermi energy, as well as, the increasing of the degeneracy of the
energy states with consequences on the transport properties of the
system.
\bigskip
\begin{figure}
{%
\includegraphics*[width=6.5cm,height=5.0cm]{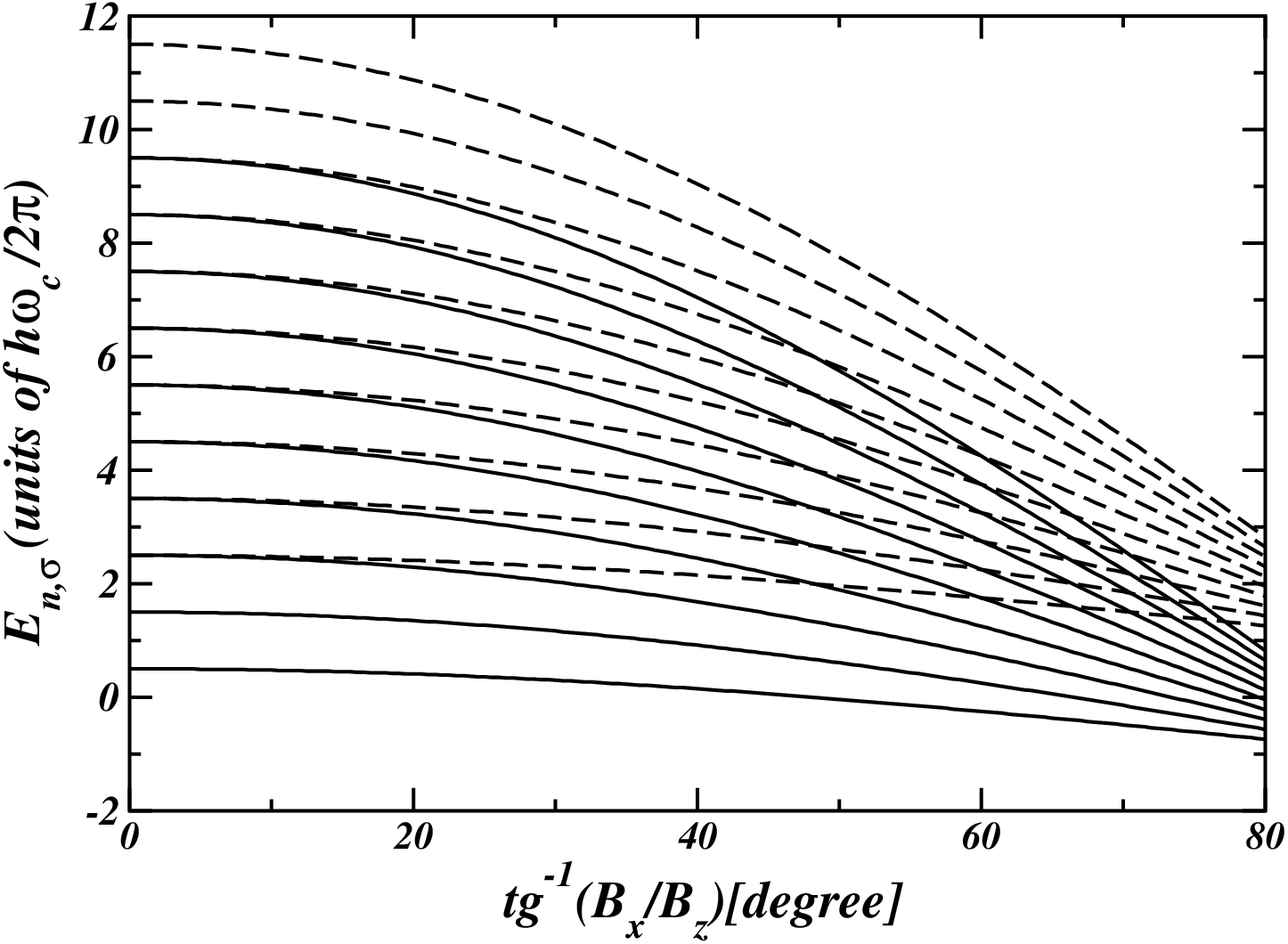}}
\caption{The energy of one-electron states with $ n
\leq 10$ against the angle between the magnetic field and the
$z$-direction. The straight (dashed) lines are for $E_{n,\sigma _-}$
($E_{n,\sigma _+}$), and n=1 is the line at the bottom.}
\label{landaulev}
\end{figure}
\bigskip
\section{\label{section 3}the Fermi Energy}

As mentioned before, a moderate or low electronic density allows
us to use the one particle approximation to obtain information
about the system, and then, a standard calculation can be used to
obtain the Fermi energy of the system. The density of state is
given by $ D(E) = \sum\limits_{nk_x \sigma } {\delta (E -
E_n^\sigma  )}$ and we may assume a Gaussian broadening of width
$\Gamma$ to write $D(E)$ as

\begin{equation}
D(E) = \frac{{S_0 }}{{(2\pi )^{3/2} }}\sum\limits_{n\sigma }
{\frac{{e^{ - (E - E_n^\sigma  )^2 /2\Gamma ^2 } }}{{l_c^2 \Gamma
}}}.
 \label{density}
\end{equation}
Here $S_0=L_xL_y$ ($L_i$ is the length of the system along the
$i$-direction), and we assume that the Gaussian has the same width
for all energy levels $E_n^\sigma$ (we use $\mu_e = 0.05$ $m_e$ and
$\Gamma = 0.5$ $meV$ in all numerical results presented here). Since
the density $n_s$ of particles (electrons) is given by
\begin{equation}
n_s = \mathop {\lim }\limits_{T \to 0} \frac{1}{{S_0 }}\int_{ -
\infty }^\infty {f(E)D(E)dE} \label{density1}
\end{equation}
with $ f(\varepsilon ) = \left[e^{(\varepsilon  - E_F )/k_B T}  + 1\right]^{
- 1}$ the Fermi-Dirac function, a straight calculation gives
\begin{equation}
n_s - \frac{1}{4\pi l_c^2}\sum\limits_{n\sigma } {\left[
{\textit{erf}\left( {\frac{{E_n^\sigma  }}{{\sqrt 2 \Gamma }}} \right) +
\textit{erf}\left( {\frac{{E_F  - E_n^\sigma  }}{{\sqrt 2 \Gamma }}}
\right)} \right]}  = 0 \label{eqfermi}
\end{equation}
where $\textit{erf}\,(\xi ) = \frac{2}{\sqrt{\pi}}\int_0^\xi {e^{
- t^2 } dt}$ is the error function. For a given electronic density
the Fermi energy is the value of $E_F$ that satisfies the
Eq.(\ref{eqfermi}).

In Figures 2a and 2b we show the behavior of the Fermi energy
against the angle between the dc magnetic field and the
$z$-direction for $n_s= 10^{11}cm^{-2}$ and $10^{12} cm^{-2}$
respectively. The oscillations observed are a consequence of
inversions in the sequence of the energy states, since all of them
converge to $\pm \hbar \omega_c/2$ when $B_z/B_x$ is very small.
As the Fermi energy is intimately related to the transport
properties of the system, this behavior has a strong influence on
this property. In Figure 3, the direction of the magnetic field is
fixed and we let its intensity to vary. The jumps (discontinuity)
observed are the result of the increasing of the splitting of the
states energy.

\begin{figure}
{%
\includegraphics*[width=6.5cm,height=5.0cm]{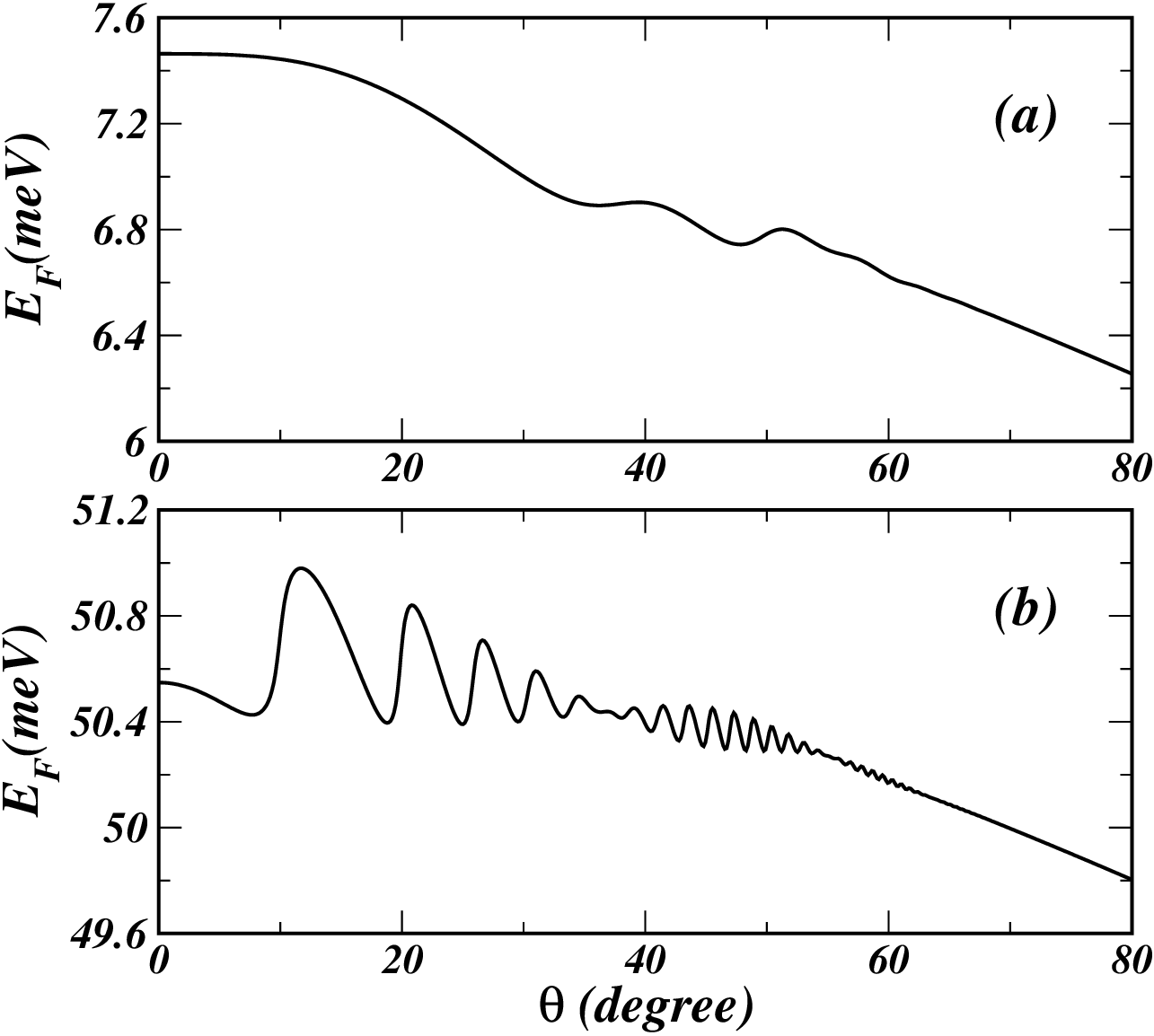}}
\caption{Fermi Energy of a 2DEG as a
function of the angle between the dc magnetic field of $10 kG$ and
the $z$-direction for two specific values of the electron density; (a)$n_s=10^{11} cm^{-2}$, (b)$n_s=10^{12} cm^{-2}$.}\label{fig2}
\end{figure}

\bigskip

\begin{figure}
{%
\includegraphics*[width=6.5cm,height=5.0cm]{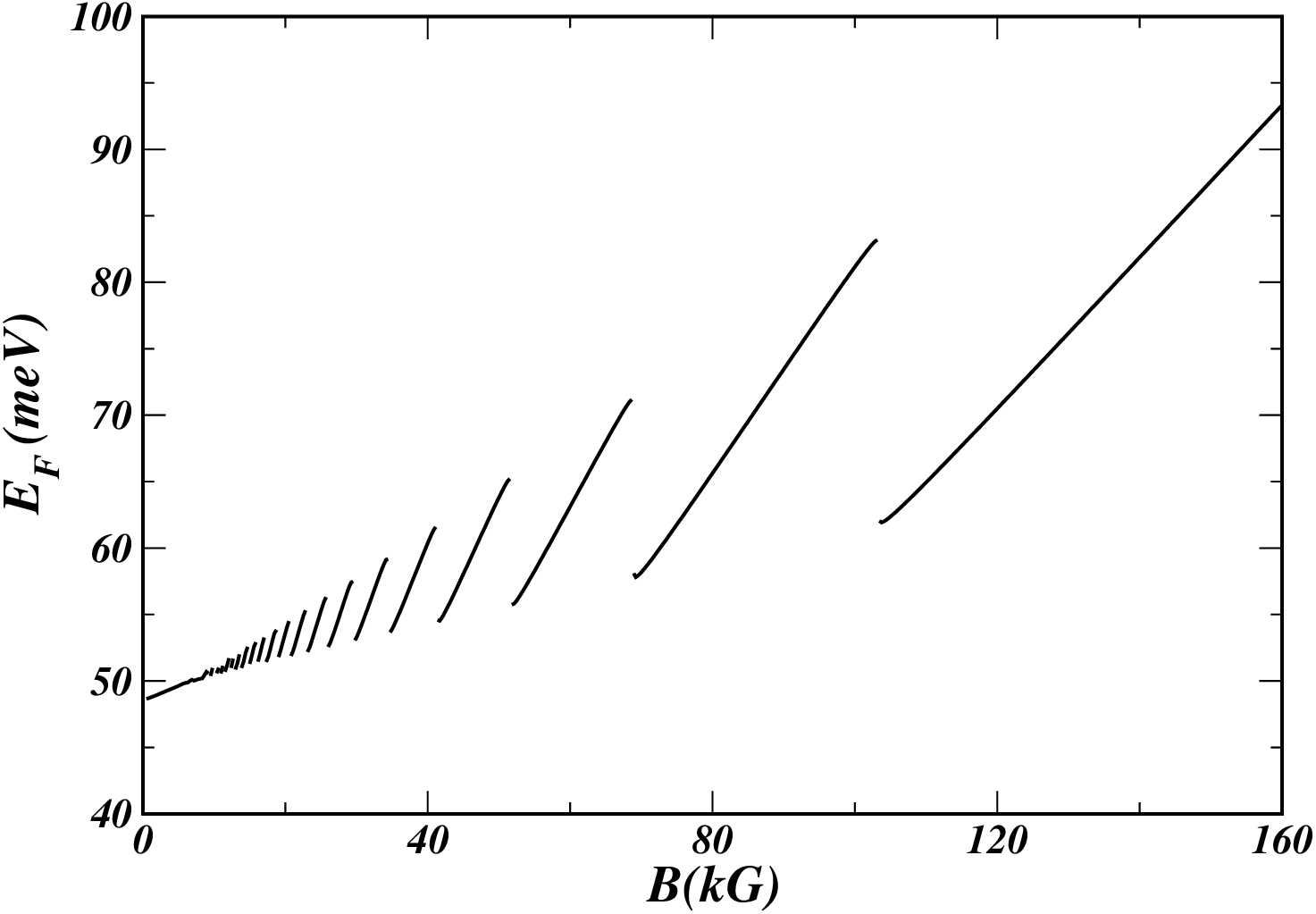}}
\caption{Fermi Energy of a 2DEG of density $10^{12} cm^{-2}$ as a
function of the intensity of a dc magnetic field applied parallel
to the $z$-direction.}

\label{fig3}
\end{figure}
%

\section{\label{section 4}the hall conductivity}
When an electric field produces a carrier current in the presence
of a non collinear magnetic field, there is always a transversal
current as a result of the magnetic field. This is the Hall effect. Since the conductivity observed experimentally incorporates several properties of the sample, this effect is used as a tool to study several physical properties and to characterize several materials.

Here, the Hall conductivity is used to estimate the influence of a
tilt dc magnetic field applied on the 2DEG. We use the results
obtained in Sections II and III to calculate the Hall
conductivity, which (within the one particle approximation) is
given by (see for example Refs. 4):

\begin{eqnarray}
\sigma_{xy}&=& \frac{2i \hbar e^2}{S_0} \sum_{\zeta,
\zeta'}f(E_{\zeta} ) [1-f(E_{\zeta'
})] \times \nonumber \\
& & \left[\frac{1-e^{(E_{\zeta}-E_{\zeta'
})/{K_BT}}}{(E_{\zeta}-E_{\zeta' })^2}\right]\langle\zeta |\widehat{v}_x | \zeta' \rangle
\langle \zeta'| \widehat{v}_y | \zeta\rangle \label{lab8}
\end{eqnarray}
with $\left| {\zeta} \right\rangle$ $\neq$ $\left| {\zeta'}
\right\rangle$, where $\left| {\zeta} \right\rangle$ is the state
$\left| {n,k_x,\sigma} \right\rangle$.

In the above equation, $S_0$ is the area of the system (we use an
unitary area in our numerical calculation), $f(\varepsilon)$ is
the Fermi-Dirac function and

\begin{eqnarray}
\left\langle\zeta\right|\hat v_x
\left| {\zeta '} \right\rangle &=& -\frac{1}{\sqrt 2 }\left[ \sqrt {n' + 1} \delta _{n,n' + 1} + \sqrt {n'} \delta _{n,n' - 1}  \right]\times\nonumber\\
& & {\omega _c l_c \cos
{\theta }}\delta _{k_x ,k_x' } \delta
_{\sigma ,\sigma '}
\end{eqnarray}
and
\begin{eqnarray}
\left\langle \zeta ' \right|\hat v_y \left| {\zeta } \right\rangle
&=&  \frac{i}{\sqrt 2 } \left[ {\sqrt {n + 1} \delta _{n,n'- 1}
-\sqrt {n} \delta _{n,n' + 1} } \right]\times\nonumber\\
& &{\omega _c l_c \cos
{\theta}}\delta _{k_x ,k_x'} \delta
_{\sigma ,\sigma '}
\end{eqnarray}
are the matrix element of the $x$ and $y$
components of the velocity operator .

Numerical results are depicted in Figures 4 and 5 for $T=0.4K$.

\begin{figure}

{%
\includegraphics*[width=6.5cm,height=5.0cm]{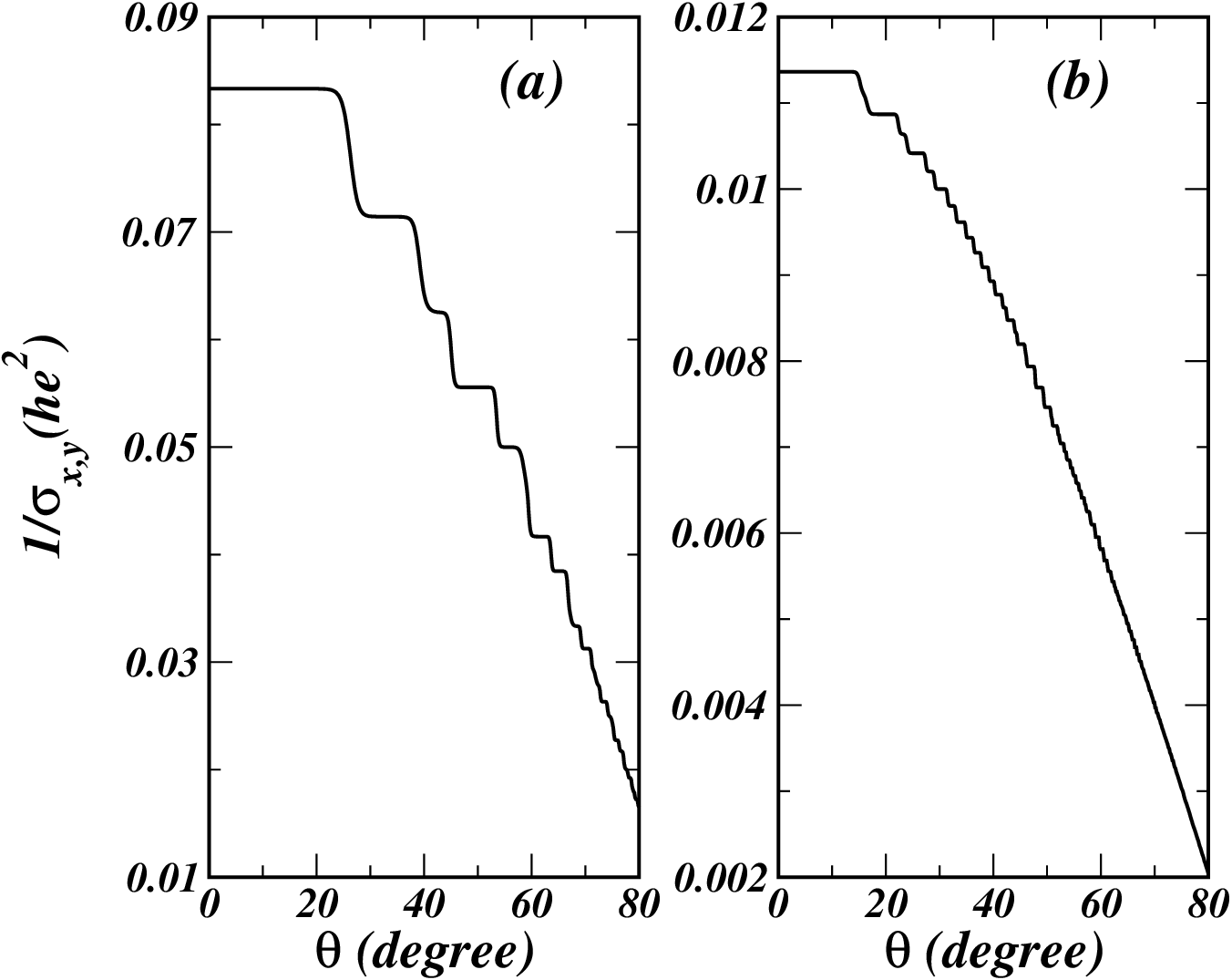}}
\caption{Hall conductivity of a 2DEG as a function of the angle
between a $10 kG$ dc magnetic field and the $z$-direction; a)
density $n_s=10^{11} cm^{-2}$; b) density $n_s=10^{12} cm^{-2}$.}
\label{fig4}
\end{figure}
\begin{figure}

{%
\includegraphics*[width=6.5cm,height=5.0cm]{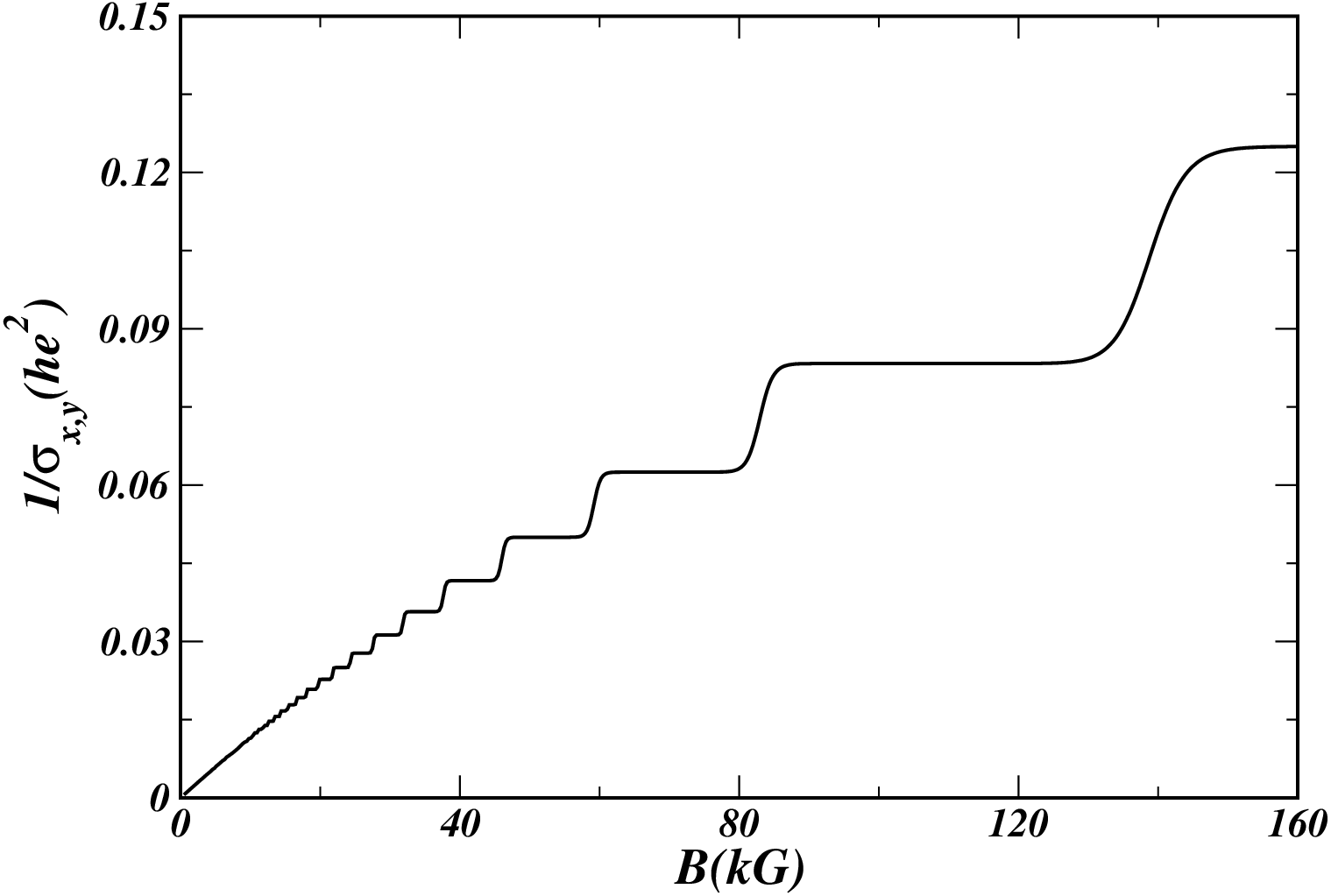}}
\caption{Hall conductivity of a 2DEG of density $10^{12} cm^{-2}$
as a function of the intensity of a dc magnetic field applied
parallel to the $z$-direction. } \label{fig5}
\end{figure}

\section{\label{section 5}final comments and conclusions}
In Figure 2 we show the Fermi energy as a function of the angle
between the magnetic field and the $z$-direction. The behavior of
the Fermi energy is a direct consequence of the behavior of the
energy of the eigenstates depicted in Figure 1. It is interesting
to point out that when $\theta $ increases, the energies of all
states go to $\pm \hbar \omega_{c}/2$ and, as a consequence, there
is an increasing of the number of states per unit of energy. The
increasing of the electronic density enhances this behavior
because the electrons are ``accommodated" in a larger number of
states and then there will be a bigger mix of populated states.

Also, for finite values of $\Gamma$, there will be a mix of the
Gaussian associated with the neighbor energy levels. When the
magnetic field increases the degeneracy of the states also
increases, and there will be values of $B(\Gamma)$ where these
states are ``disconnected" resulting (at these points) in a
singular behavior for the Fermi energy. It should be said that if
$\Gamma$ is changed, the shape of the curves are different but the
general behavior is preserved.

For electronic density that has the Fermi energy located at the
Landau energy larger than a given $E_n$ but much smaller than
$E_{n+1}$ the increasing of its value with the external field has
a smaller derivative than the case that the Fermi energy is bellow
$E_n$, but much bigger than $E_{n-1}$. This is the main reason to
the oscillatory behavior observed when the field is modified. The
Hall conductivity also reproduce this result, and it has well
defined ``terraces" with length related to the ``distance" of two
consecutive minima (or discontinuity) of the Fermi energy.

This behavior is better noticed when the density of electrons
increases and stronger oscillations are observed for the Fermi
energy (see Figures 2 and 4). However, the length of the terrace
decreases because the distance between the local minima is
smaller.

Our results show that the conductivity decreases when the magnetic
field increases (see Figure 5) while the parallel component of the
field provides a way to increase it. In both cases we can see well
defined plateaus with length related to the Fermi energy behavior.

In summary, our calculation shows that the Fermi energy is
strongly dependent on the spin states, which is controlled by the
external dc field. Moreover, any property that depends on the
Fermi-Dirac distribution, (which is dependent on the Fermi energy)
will have the influence of these spin states. We use the Hall
conductivity to show this fact. We are now working on a more
complex system to demonstrate that the spin states may be used to
control the transport properties of selected materials. This might
be the reason for the behavior of the resistivity recently
observed by K. Vakili et al.\cite{ref9,ref10}.
\begin{acknowledgments}

ACAR and NSA thank the CNPq and the Funda\c{c}\~{a}o Cearense de
Apoio a Pesquisa (FUNCAP) for the financial support. GAF and RNCF
are partially supported by the Brazilian National Research
Council (CNPq).

\end{acknowledgments}

\end{document}